\documentclass[sigconf,nonacm,natbib=true]{acmart}
\AtBeginDocument{%
  }

\usepackage{makecell}

\makeatletter
\newcommand{\acmrightssize}{\fontsize{8}{9.5}\selectfont}

\newcommand{\firstpagerights}[1]{%
  \begingroup
    \renewcommand\thefootnote{}%
    \footnotetext{%
      \acmrightssize
      \raggedright
      \setlength{\parskip}{0pt}%
      \setlength{\parindent}{0pt}%
      #1%
    }%
    \addtocounter{footnote}{0}%
  \endgroup
}
\makeatother

\begin{document}

\title{From SERPs to Agents: A Platform for Comparative Studies of Information Interaction}

\author{Saber Zerhoudi}
\orcid{0000-0003-2259-0462}
\affiliation{%
  \institution{University of Passau}
  \city{Passau}
  \country{Germany}
}
\email{saber.zerhoudi@uni-passau.de}

\author{Michael Granitzer}
\orcid{0000-0003-3566-5507}
\affiliation{%
  \institution{University of Passau}
  \city{Passau}
  \country{Germany}
}
\affiliation{%
  \institution{Interdisciplinary Transformation University Austria}
  \city{Linz}
  \country{Austria}
}
\email{michael.granitzer@uni-passau.de}

\renewcommand{\shortauthors}{S. Zerhoudi et al.}

\begin{abstract}
The diversification of information access systems, from RAG to autonomous agents, creates a critical need for comparative user studies. However, the technical overhead to deploy and manage these distinct systems is a major barrier. We present UXLab~\footnotemark[1], an open-source system for web-based user studies that addresses this challenge. Its core is a web-based dashboard enabling the complete, no-code configuration of complex experimental designs. Researchers can visually manage the full study, from recruitment to comparing backends like traditional search, vector databases, and LLMs. We demonstrate UXLab's value via a micro case study comparing user behavior with RAG versus an autonomous agent. UXLab allows researchers to focus on experimental design and analysis, supporting future multi-modal interaction research.
\end{abstract}

% --- ACM CCS Concepts (put after the abstract, before \maketitle) ---


\maketitle
\firstpagerights{%
  © ACM, 2026. This is the author's version of the work.\\
  The definitive version was published in:
  \emph{Proceedings of the 2026 ACM SIGIR Conference on Human Information Interaction and Retrieval (CHIIR '26),
  March 22--26, 2026, Seattle, WA, USA}.\\
  DOI: \url{https://doi.org/10.1145/3786304.3787948}
}

\footnotetext[1]{Github repository: \url{https://github.com/searchsim-org/uxlab}}
\footnotetext[1]{Website: \url{https://uxlab.searchsim.org}}

\begin{figure}[t]
  \centering
  \includegraphics[width=1\linewidth]{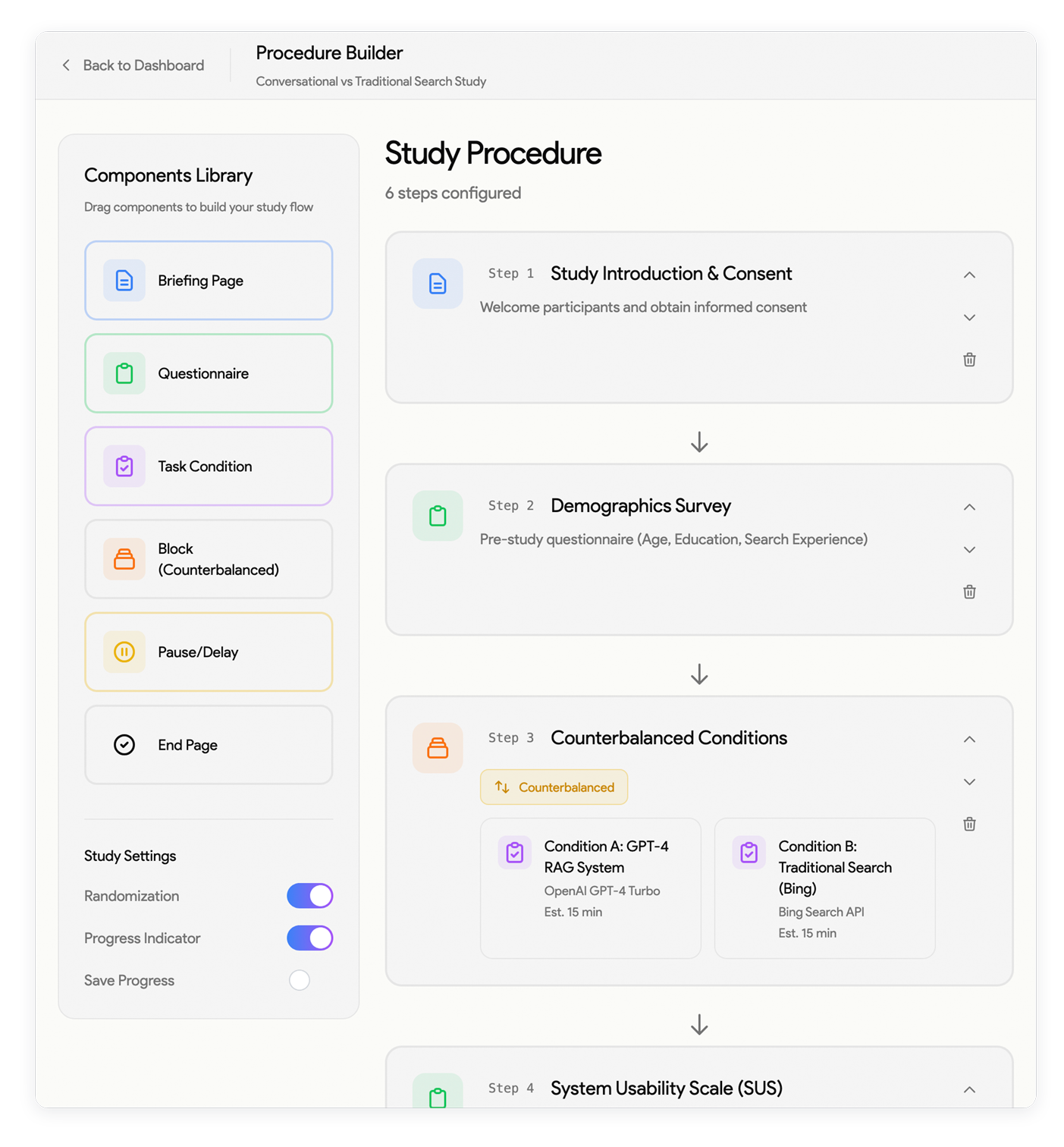}
  \caption{Overview of the UXLab Experimenter Dashboard.}
  \label{fig:overview}
  % \vspace{-18pt}
\end{figure}

\section{Introduction}
Understanding how user strategy, cognitive load, and trust adapt when shifting from traditional search to RAG-based systems~\cite{lewis2021retrievalaugmentedgenerationknowledgeintensivenlp,gao2024retrievalaugmentedgenerationlargelanguage} or autonomous agents~\cite{yao2023reactsynergizingreasoningacting} is a critical HCI research question~\cite{hancock2020ai,liao2023aitransparencyagellms}. However, executing such a comparative study presents a substantial engineering barrier. A researcher must first build, deploy, and integrate these three distinct systems into a single, cohesive experimental framework. This framework must also handle participant management, study logic, and high-fidelity logging. This technical overhead forces a shift in focus from human-computer interaction to backend software engineering~\cite{Kelly09}.

This situation underscores the need for a standardized, reusable infrastructure. While foundational IR libraries like Pyserini~\cite{lin2021pyserinieasytousepythontoolkit} are essential for algorithms, they do not function as end-to-end experimental platforms. General-purpose experiment builders~\cite{zamani2019macawextensibleconversationalinformation,Fr_be_2023}, in turn, lack the capability to manage the complex backend configurations required for modern IR and AI research.

To address this gap, we introduce \textbf{UXLab}, an open-source, modular software system. UXLab is an end-to-end, no-code workbench that allows researchers to rapidly design, deploy, and manage complex, comparative user studies of information access systems.

The primary contribution is the \textbf{Experimenter Dashboard}. This web interface provides no-code control over a flexible, four-part architecture: (1) a core \textbf{Backend} for study logic, (2) the \textbf{Participant Interface}, (3) the dashboard itself for configuration, and (4) modular \textbf{Service Connectors} for integrating any external or local search, RAG, or agentic system.

Our system enables researchers to configure and deploy comparative studies in hours, a process that traditionally takes months. We demonstrate this utility through a focused case study: a within-subjects (N=8) comparison of a RAG system versus an agentic system for complex tasks. The framework enhances reproducibility by allowing researchers to share complete experimental setups as a single file. To situate our contribution, we first present the experimental workflow (Sec.~\ref{sec:workflow}) and then detail the system's architecture (Sec.~\ref{sec:architecture}). The framework is open-source and available online~\footnotemark[1].

\section{Related Work}
Building robust, reusable infrastructure for human-computer interaction studies is a long-standing challenge. The contribution of UXLab is best understood by situating it within the landscape of existing tools, which we divide into three main categories.

\subsection{General Experiment Frameworks}
General-purpose frameworks like jsPsych~\cite{LeeuwGL23} or StudyAlign~\cite{Lehmann_2025} effectively manage user study logistics, such as participant flow. However, these systems are content-agnostic by design and do not provide the core information access system (e.g., a search engine or RAG pipeline) to be evaluated. Consequently, researchers must build, deploy, and maintain a separate backend prototype before they can conduct the experiment.

%``quoted text''
\subsection{Information Retrieval and NLP Toolkits}
Code-level toolkits like Pyserini~\cite{lin2021pyserinieasytousepythontoolkit}, LangChain~\cite{langchain2023}, and the Hugging Face libraries~\cite{wolf-etal-2020-transformers} provide essential components for building retrieval and agentic systems. They are not, however, end-to-end experimental platforms. A researcher must still engineer the complete application infrastructure, including the backend server, frontend UI, and logging database. This reliance on significant engineering expertise limits adoption by the broader community.

\subsection{Specialized Research Platforms}
A third category consists of specialized platforms focused on a single interaction modality, such as Chatty Goose for conversational search~\cite{Zhang2021ChattyGoose}. While these tools are effective for research \emph{within} their paradigm (e.g., comparing two RAG models), their specialized design makes them unsuitable for comparative studies \emph{between} paradigms, such as evaluating a conversational agent against a traditional SERP in a unified experiment.

\subsection{The UXLab Niche}
UXLab is designed to fill the gap between these categories. It operates as an integrated, end-to-end system, similar to StudyAlign~\cite{Lehmann_2025}, yet remains specialized for information interaction research, much like Chatty Goose~\cite{Zhang2021ChattyGoose}. Its primary contribution is the Experimenter Dashboard, a no-code interface that connects high-level experimental design to low-level backend configuration.

Unlike code libraries, UXLab is a complete, deployable platform. Unlike general-purpose tools, it treats IR and AI backends as first-class, configurable features. Furthermore, unlike other specialized platforms, it is comparative by design, allowing researchers to treat traditional search, RAG, and agentic systems as interchangeable ``conditions'' within a single, unified experimental framework.

\section{The UXLab System Architecture}
The UXLab system is designed to decouple researcher configuration from participant interaction. Its architecture comprises four primary components: the Backend, the Experimenter Dashboard, the Participant Interface, and the Service Connectors. Before detailing these components, we first outline the research workflow UXLab is designed to support.

\subsection{The UXLab Research Workflow}
\label{sec:workflow}
UXLab is designed to simplify the technical overhead of web-based experiments. 

The platform's workflow separates conceptual tasks, performed by the researcher, from technical automation tasks. The researcher's conceptual work involves:

\begin{itemize}
    \item \textbf{Formulating the Research Question:} Defining the hypothesis and the experimental design.
    \item \textbf{Preparing Backends:} Ensuring external services are running (e.g., a local Ollama server or Lucene instance) or obtaining necessary API keys (e.g., for OpenAI).
\end{itemize}

\noindent UXLab then automates the complete technical setup and execution:
\begin{itemize}
    \item \textbf{Create Study:} The researcher initiates a new study using the web dashboard.
    \item \textbf{Configure Backends:} Researchers select their backends in the UI (e.g., \texttt{OpenAI}, \texttt{Local URL}) and provide the required keys or addresses.
    \item \textbf{Build Procedure:} The researcher uses a visual editor to construct the study flow, arranging tasks, questionnaires, and backend conditions.
    \item \textbf{Deploy \& Recruit:} UXLab generates a single, shareable URL for participant distribution (e.g., via Prolific~\cite{prolific2025}).
    \item \textbf{Monitor \& Export Data:} The platform provides real-time progress monitoring and allows the researcher to export all interaction logs and responses as a single CSV file.
\end{itemize}

\subsection{System Architecture Overview}
\label{sec:architecture}
\begin{figure}[ht!]
  \centering
  \includegraphics[width=.9\linewidth]{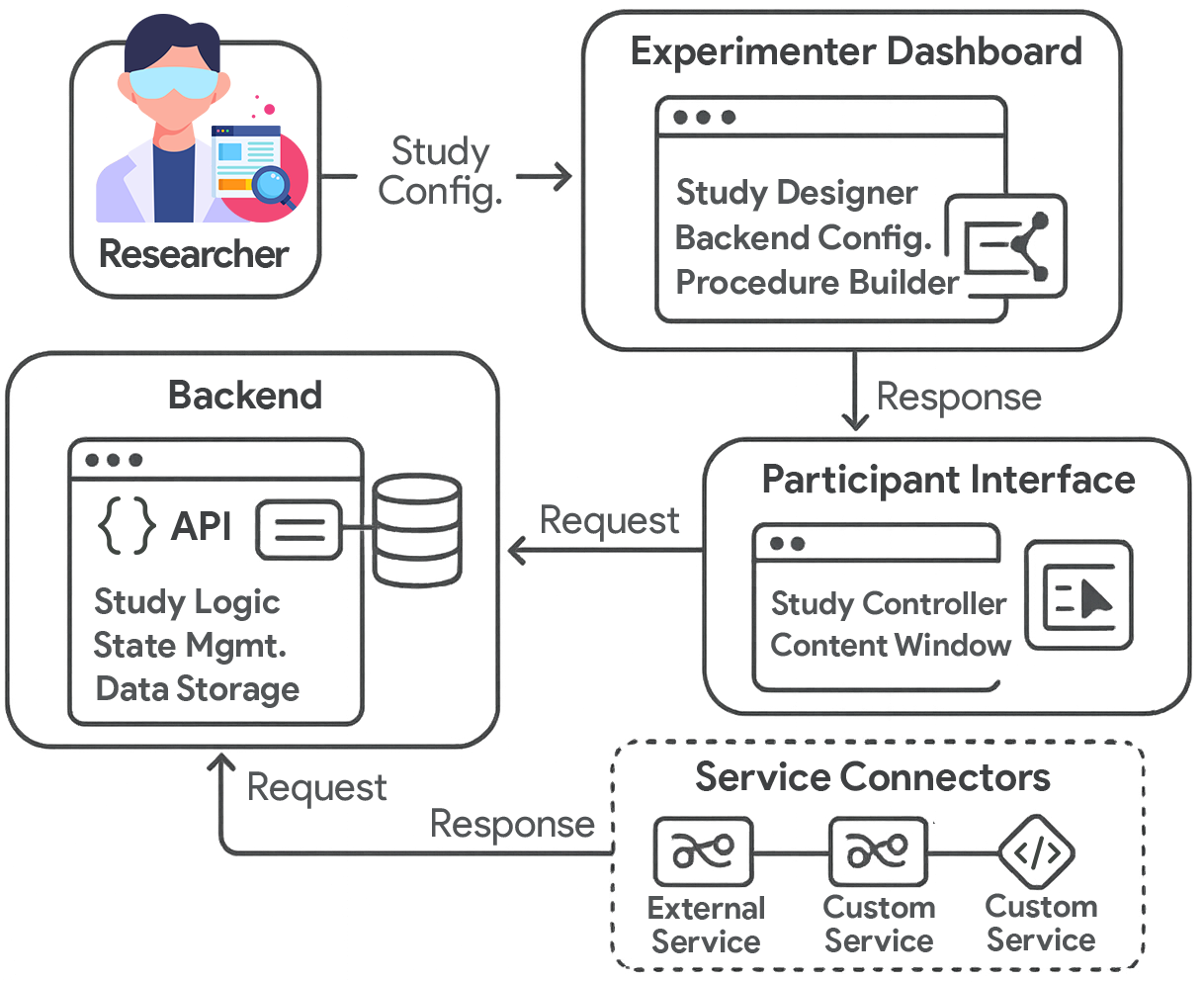}
  \caption{UXLab system architecture. The Experimenter Dashboard configures the Backend, which routes all requests from the Participant Interface via Service Connectors to the appropriate external or local services.}
  \label{fig:arch_overview}
  % \vspace{-8pt}
\end{figure}

This workflow is enabled by the system's four-part architecture, which uses a client-server model (Fig.~\ref{fig:arch_overview}).

\subsubsection{The Core Backend}

The Backend is the system's central controller, built with FastAPI (Python 3.10+). Its responsibilities include:
\begin{itemize}
    \item \textbf{API Provision:} Offering a secure REST API for the dashboard and participant interface.
    \item \textbf{Study Logic:} Managing experimental logic, such as participant counterbalancing (e.g., Latin square) and assignment.
    \item \textbf{State Management:} Tracking each participant's progress.
    \item \textbf{Data Persistence:} Storing all study configurations, logs, and responses in a PostgreSQL database.
\end{itemize}

\subsubsection{The Experimenter Dashboard (Control Panel)}

The web application provides researchers with no-code control over the experiment lifecycle. Its key modules are:
\begin{itemize}
    \item \textbf{Study Designer:} A visual interface for study creation, defining experimental groups, and managing recruitment.
    \item \textbf{Backend Configurator:} A control panel to define and credential the systems under study (e.g., ``Bing API'', ``Ollama''). This permits a single participant interface to be connected to different backends for comparative A/B testing.
    \item \textbf{Procedure \& Questionnaire Builder:} A visual editor to define the study flow. Researchers can assemble text pages, external questionnaires, and ``Task'' elements that load a configured backend.
    \item \textbf{Participant Management:} Supports Prolific and MTurk~\cite{Crowston2012AMT} integration for ID assignment and code redirection.
    \item \textbf{Data Export:} A function to export all timestamped logs and questionnaire data for analysis.
\end{itemize}

\subsubsection{The Participant Interface (Study Frontend)}

This is the minimalistic, participant-facing web application. It is designed for robustness and consists of two parts:
\begin{itemize}
    \item \textbf{The Study Controller:} A persistent frame controlled by the backend. It shows the task briefing and the ``Next'' button, managing the participant's progression.
    \item \textbf{The Content Window:} An embedded \texttt{iframe} showing the current element (e.g., text, questionnaire, or the interactive task prototype).
\end{itemize}
This separation is key: it lets UXLab manage any web-based prototype without requiring modification of the prototype itself.

\subsubsection{The Service Connectors (The ``Library'')}

This is a modular layer of Python classes on the backend. They act as a standard interface between UXLab and external services. A request from the participant (e.g., a search) is sent to the backend. The backend routes this to the correct Service Connector (e.g., ``\texttt{OllamaConnector}''). The connector translates this standard request into the specific format for the external API, sends it, and translates the response back. This design is the key to UXLab's flexibility:
\begin{itemize}
    \item \textbf{Compare Systems:} Researchers can run A/B tests (e.g., Bing vs. Google) with no frontend code changes.
    \item \textbf{Extend the Platform:} Support for a new, custom-built system can be implemented simply by writing a new Connector class that adheres to a simple API standard.
\end{itemize}

\section{Supported Experimental Designs}
UXLab's architecture, which separates the Experimenter Dashboard from the Participant Interface, was designed to directly support common experimental methodologies in CHI and IR. Researchers can visually construct these study designs without code using the system's ``Procedure Builder'' (Fig.~\ref{fig:overview}) and ``Backend Configurator''.

\subsection{Between-Subject Designs}
This design is used to compare distinct groups of participants, for example, comparing a ``Traditional'' interface (Group A) against an ``Agentic'' one (Group B).

\textbf{UXLab Implementation:} The researcher uses the ``Duplicate Study'' feature. They first configure the complete study for Group A, setting its ``Condition'' to use a specific backend (e.g., a standard search API). They then duplicate this study, name it ``Group B,'' and simply edit the ``Condition'' element in this new version to point to a different backend (e.g., an agentic endpoint). Participants are then randomly assigned one of the two study links.

\subsection{Within-Subject Designs}
This design, where each participant experiences all conditions, is powerful for comparative studies. A participant might be asked to test both a RAG and an Agentic interface.

\textbf{UXLab Implementation:} This is a core function of the Procedure Builder. The researcher creates two ``Condition'' elements (e.g., ``Condition RAG'' and ``Condition Agentic''). They are then placed into a ``Block'' element, where the researcher checks a single box: ``Counterbalance''. The UXLab backend automatically manages the assignment logic, ensuring participants are routed through a randomized order (e.g., A-B or B-A) without manual configuration.

\subsection{Interrupted Time-Series Studies}
UXLab also supports time-delayed experiments, such as studying how user behavior changes after a period of adaptation. This allows for examining personalization or other long-term effects.

\textbf{UXLab Implementation:} The researcher adds a ``Pause'' element to the procedure flow.

This element can be configured to be time-based (e.g., ``Continue in 3 days'') or manually controlled (e.g., ``Wait for experimenter approval''). This permits researchers to conduct an intervention, such as re-training a model with the participant's data, before allowing them to proceed, enabling sophisticated, multi-day study designs.

\begin{table*}[ht!]
\centering

\begin{minipage}[t]{0.5\textwidth}
\centering
\small
\caption{Comparison of behavioral metrics (N=8 participants).}
\label{tab:behavioral}
\begin{tabular}{l c c c c}
\toprule
\textbf{Metric} &
\makecell{\textbf{RAG}\\(Mean, SD)} &
\makecell{\textbf{Agentic}\\(Mean, SD)} &
\textbf{t(23)} &
\textbf{p-value} \\
\midrule
Time (sec) & 310.5 ($\pm$ 95.2) & \textbf{245.1} ($\pm$ 71.3) & 3.88 & $<0.001$ \\
User Follow-ups & \textbf{4.8} ($\pm$ 2.1) & 2.1 ($\pm$ 1.4) & 5.91 & $<0.001$ \\
Initial Query (chars) & 58.3 ($\pm$ 22.9) & \textbf{81.7} ($\pm$ 30.5) & -3.12 & $<0.01$ \\
\bottomrule
\end{tabular}
\end{minipage}
\hfill%
\begin{minipage}[t]{0.45\textwidth}
\centering
\small
\caption{Mean user-reported ratings (1–5 Likert scale).}
\label{tab:satisfaction}
\begin{tabular}{p{0.60\linewidth} c c}
\toprule
\textbf{Statement} & \textbf{RAG} & \textbf{Agentic} \\
\midrule
``I am satisfied with the final answer.'' & 3.6 & \textbf{4.4} \\
``The task was mentally demanding.'' & \textbf{4.1} & 2.5 \\
``I trusted the system to complete the task.'' & 3.3 & \textbf{4.2} \\
``I felt in control of the search process.'' & \textbf{4.5} & 3.1 \\
\bottomrule
\end{tabular}
\end{minipage}

\end{table*}

\section{Case Study: RAG vs. Agentic Search}

To demonstrate UXLab's utility as an experimental tool, we conducted a focused, within-subjects micro study. This case study serves as a \textbf{proof-of-concept}, illustrating how the platform can be used to investigate a research question that is otherwise technically difficult to set up.

\subsection{Experimental Goal and Design}
The study's goal was to compare user behavior, satisfaction, and cognitive load when using a standard RAG system versus a multi-step Agentic system for complex, multi-faceted tasks.

We recruited \textbf{8 participants} from Prolific platform. We used a within-subject design where each participant completed two complex search tasks (e.g., planning a 3-day itinerary). The task order and the order of the two system conditions were counterbalanced. Participants completed a post-questionnaire after each task.

\subsection{UXLab Configuration}
The entire study was configured and deployed using the \textbf{Experimenter Dashboard} in less than two hours.

\begin{enumerate}
    \item \textbf{Backend Configuration:} In the ``Backend Configurator,'' we created two distinct conditions. For \textbf{Condition 1 (RAG)}, we selected the ``OpenAI'' connector and provided a prompt template for single-step retrieval-augmented generation. For \textbf{Condition 2 (Agentic)}, we also used the ``OpenAI'' connector but provided a different prompt template and enabled ``Agentic Mode'' to allow for autonomous, multi-step plan execution via its search tool.
    \item \textbf{Procedure Building:} We used the ``Procedure Builder'' to create the study flow, which included consent pages, a counterbalanced block assigning the two conditions, and post-task questionnaires.
    \item \textbf{Deployment:} From the ``Participant Management'' tab, a single study link was generated and posted to Prolific. The UXLab backend automatically managed all participant assignment and counterbalancing.
\end{enumerate}

\subsection{Analysis of Collected Session Data}
All behavioral and questionnaire data were collected by the UXLab backend and exported as a single CSV file. We used paired t-tests for statistical comparisons.

\paragraph*{Behavioral Analysis}
Analysis of interaction metrics (Table \ref{tab:behavioral}) showed a shift in user behavior. Users completed tasks faster ($p < .001$) and issued fewer follow-up queries ($p < .001$) when using the Agentic system. Users adapted their interaction style by submitting longer, more descriptive initial ``task delegations'' ($p < .01$) instead of shorter, single-question ``queries.''

\paragraph*{User Satisfaction and Trust}
Post-task questionnaire data (Table~\ref{tab:satisfaction}) revealed that the Agentic system scored higher on satisfaction and trust, while reducing cognitive load. This improvement, however, resulted in lower scores for perceived user control, which highlights a critical \textbf{Trust versus Control trade-off} inherent in agentic system design. This case study successfully validates the platform's core utility, demonstrating its ability to be deployed for novel, comparative research that would otherwise be technically difficult.

\section{Discussion}

The case study demonstrates UXLab's effectiveness for novel comparative research. The system's value, however, is in the methodological power it provides. By standardizing experiment management, UXLab reduces the time from research question to active study. Its ``Service Connector'' architecture treats backends as modular components, enabling robust comparisons \emph{between} paradigms (e.g., SERP vs. RAG) as easily as \emph{within} them. It ensures methodological robustness through tested, reusable logic for counterbalancing and standardized, analysis-ready data logging.

\paragraph{Scope}

It is important to clarify UXLab's scope: it is \emph{not} a prototyping tool. Researchers continue to build their own functional prototypes. UXLab’s primary role is to act as the ``scaffolding'' \emph{around} these prototypes---a system to manage the experimental procedure, connect components, control participant flow, and log all data. Extensibility is achieved via the Service Connector model, where integrating a new system requires implementing only a single Python class.

\paragraph{Transparency and Reproducibility}
UXLab directly addresses the need for greater research transparency. Reproducibility for complex AI-based studies requires sharing the exact configuration and experimental design. The Experimenter Dashboard features a ``one-click'' export that saves the entire study as a single JSON file. This file can be shared with a publication, allowing any other researcher to import it into their own UXLab instance and replicate the study identically.

\section{Conclusion and Future Work}
This paper introduced \textbf{UXLab}, an open-source system for conducting web-based user studies on information interaction. We identified a critical gap in research infrastructure: the prohibitive engineering effort required for comparative studies across traditional, RAG, and agentic search paradigms.

UXLab solves this by providing four core components: (1) a robust Backend, (2) an Experimenter Dashboard for no-code configuration, (3) a participant-facing Interface for procedural control, and (4) a modular Service Connector library. We demonstrated the system's utility through a case study comparing RAG and Agentic systems, an experiment configured in hours, not months.

By abstracting the engineering complexity, UXLab enables researchers to focus on designing experiments and understanding human-computer interaction. We plan to expand the library of Service Connectors and publish UXLab as an open-source project, inviting the community to contribute to this shared, versatile tool for the field.

% ---------- Bibliography ----------
\bibliographystyle{ACM-Reference-Format}
\bibliography{sample-base}

@String{Computing = "Computing" }

@String{Springer = "Springer-Verlag" }

@misc{liao2023aitransparencyagellms,
      title={AI Transparency in the Age of LLMs: A Human-Centered Research Roadmap}, 
      author={Q. Vera Liao and Jennifer Wortman Vaughan},
      year={2023},
      eprint={2306.01941},
      archivePrefix={arXiv},
      primaryClass={cs.HC},
      url={https://arxiv.org/abs/2306.01941}, 
}

@article{hancock2020ai,
  title={AI-mediated communication: Definition, research agenda, and ethical considerations},
  author={Hancock, Jeffrey T and Naaman, Mor and Levy, Karen},
  journal={Journal of Computer-Mediated Communication},
  volume={25},
  number={1},
  pages={89--100},
  year={2020},
  publisher={Oxford University Press}
}

@misc{lewis2021retrievalaugmentedgenerationknowledgeintensivenlp,
      title={Retrieval-Augmented Generation for Knowledge-Intensive NLP Tasks}, 
      author={Patrick Lewis and Ethan Perez and Aleksandra Piktus and Fabio Petroni and Vladimir Karpukhin and Naman Goyal and Heinrich Küttler and Mike Lewis and Wen-tau Yih and Tim Rocktäschel and Sebastian Riedel and Douwe Kiela},
      year={2021},
      eprint={2005.11401},
      archivePrefix={arXiv},
      primaryClass={cs.CL},
      url={https://arxiv.org/abs/2005.11401}, 
}

@misc{yao2023reactsynergizingreasoningacting,
      title={ReAct: Synergizing Reasoning and Acting in Language Models}, 
      author={Shunyu Yao and Jeffrey Zhao and Dian Yu and Nan Du and Izhak Shafran and Karthik Narasimhan and Yuan Cao},
      year={2023},
      eprint={2210.03629},
      archivePrefix={arXiv},
      primaryClass={cs.CL},
      url={https://arxiv.org/abs/2210.03629}, 
}

@article{Kelly09,
  author       = {Diane Kelly},
  title        = {Methods for Evaluating Interactive Information Retrieval Systems with
                  Users},
  journal      = {Found. Trends Inf. Retr.},
  volume       = {3},
  number       = {1-2},
  pages        = {1--224},
  year         = {2009},
  url          = {https://doi.org/10.1561/1500000012},
  doi          = {10.1561/1500000012}
}

@misc{lin2021pyserinieasytousepythontoolkit,
      title={Pyserini: An Easy-to-Use Python Toolkit to Support Replicable IR Research with Sparse and Dense Representations}, 
      author={Jimmy Lin and Xueguang Ma and Sheng-Chieh Lin and Jheng-Hong Yang and Ronak Pradeep and Rodrigo Nogueira},
      year={2021},
      eprint={2102.10073},
      archivePrefix={arXiv},
      primaryClass={cs.IR},
      url={https://arxiv.org/abs/2102.10073}, 
}

@misc{gao2024retrievalaugmentedgenerationlargelanguage,
      title={Retrieval-Augmented Generation for Large Language Models: A Survey}, 
      author={Yunfan Gao and Yun Xiong and Xinyu Gao and Kangxiang Jia and Jinliu Pan and Yuxi Bi and Yi Dai and Jiawei Sun and Meng Wang and Haofen Wang},
      year={2024},
      eprint={2312.10997},
      archivePrefix={arXiv},
      primaryClass={cs.CL},
      url={https://arxiv.org/abs/2312.10997}, 
}

@inproceedings{Fr_be_2023, 
   title={The Information Retrieval Experiment Platform},
   url={http://dx.doi.org/10.1145/3539618.3591888},
   DOI={10.1145/3539618.3591888},
   booktitle={Proceedings of the 46th International ACM SIGIR Conference on Research and Development in Information Retrieval},
   publisher={ACM},
   author={Fröbe, Maik and Reimer, Jan Heinrich and MacAvaney, Sean and Deckers, Niklas and Reich, Simon and Bevendorff, Janek and Stein, Benno and Hagen, Matthias and Potthast, Martin},
   year={2023},
   month=jul, pages={2826–2836},
   collection={SIGIR ’23} 
}

@misc{zamani2019macawextensibleconversationalinformation,
      title={Macaw: An Extensible Conversational Information Seeking Platform}, 
      author={Hamed Zamani and Nick Craswell},
      year={2019},
      eprint={1912.08904},
      archivePrefix={arXiv},
      primaryClass={cs.IR},
      url={https://arxiv.org/abs/1912.08904}, 
}

@article{Lehmann_2025,
   title={StudyAlign: A Software System for Conducting Web-Based User Studies with Functional Interactive Prototypes},
   volume={9},
   ISSN={2573-0142},
   url={http://dx.doi.org/10.1145/3733053},
   DOI={10.1145/3733053},
   number={4},
   journal={Proceedings of the ACM on Human-Computer Interaction},
   publisher={Association for Computing Machinery (ACM)},
   author={Lehmann, Florian and Buschek, Daniel},
   year={2025},
   month=jun, pages={1–26}
}

@inproceedings{Zhang2021ChattyGoose,
  author    = {E. Zhang and (others)},
  title     = {Chatty Goose: A Python Framework for Conversational Search},
  booktitle = {Proceedings of the 44th International ACM SIGIR Conference on Research and Development in Information Retrieval (SIGIR ’21)},
  year      = {2021},
  pages     = {–-},    
  doi       = {10.1145/3404835.3462782},  
  url       = {https://dl.acm.org/doi/10.1145/3404835.3462782}
}

@article{LeeuwGL23,
  author       = {Joshua R. de Leeuw and
                  Rebecca A. Gilbert and
                  Bj{\"{o}}rn Luchterhandt},
  title        = {jsPsych: Enabling an Open-Source Collaborative Ecosystem of Behavioral
                  Experiments},
  journal      = {J. Open Source Softw.},
  volume       = {8},
  number       = {87},
  pages        = {5351},
  year         = {2023},
  url          = {https://doi.org/10.21105/joss.05351},
  doi          = {10.21105/JOSS.05351}
}

@misc{langchain2023,
  title        = {LangChain: A Framework for Developing Applications Powered by Large Language Models},
  author       = {Harrison Chase and the LangChain Developers},
  howpublished = {\url{https://python.langchain.com/docs/introduction/}},
  year         = {2023}
}

@inproceedings{wolf-etal-2020-transformers,
  title     = {Transformers: State-of-the-Art Natural Language Processing},
  author    = {Thomas Wolf and Lysandre Début and Victor Sanh and Julien Chaumond and Clément Delangue and Anthony Moi and Pierric Cistac and Tim Rault and Rémi Louf and Morgan Funtowicz and Joe Davison and Sam Shleifer and Patrick von Platen and Clara Ma and Yacine Jernite and Julien Plu and Canwen Xu and Teven Le Scao and Sylvain Gugger and Mariama Drame and Quentin Lhoest and Alexander M. Rush},
  booktitle = {Proceedings of the 2020 Conference on Empirical Methods in Natural Language Processing: System Demonstrations},
  pages     = {38--45},
  year      = {2020},
  doi       = {10.18653/v1/2020.emnlp-demos.6}
}

@misc{prolific2025,
  author       = {Prolific},
  title        = {Prolific – research participant recruitment platform},
  howpublished = {\url{https://www.prolific.com/}},
  note         = {Accessed: 2025-10-30},
  year         = {2025}
}

@incollection{Crowston2012AMT,
  author    = {Kevin Crowston and Neal R. Prestopnik},
  title     = {Amazon Mechanical Turk: A Research Tool for Organizations and Information Systems Scholars},
  booktitle = {Shaping the Future of ICT Research: Methods and Approaches},
  series    = {IFIP Advances in Information and Communication Technology},
  volume    = {389},
  pages     = {210--221},
  year      = {2012},
  publisher = {Springer},
  doi       = {10.1007/978-3-642-35142-6_14}
}

\end{document}